\documentclass[9pt,twocolumn,twoside]{osajnl}

\journal{ol} 

\usepackage{siunitx}

\setboolean{shortarticle}{true}

\title{1 kW, 10 mJ, 120 fs coherently combined fiber CPA laser system}

\author[1,*,$\dagger$]{Henning Stark}
\author[1,$\dagger$]{Joachim Buldt}
\author[1]{Michael M\"uller}
\author[1,2]{Arno Klenke}
\author[1,2,3]{Jens Limpert}

\affil[1]{Institute of Applied Physics, Abbe Center of Photonics, Friedrich Schiller University Jena, Albert-Einstein-Str. 15, 07745 Jena, Germany}
\affil[2]{Helmholtz-Institute Jena, Fröbelstieg 3, 07743 Jena, Germany}
\affil[3]{Fraunhofer Institute for Applied Optics and Precision Engineering, Albert-Einstein-Str. 7, 07745 Jena, Germany}

\affil[*]{Corresponding author: lars.henning.stark@uni-jena.de}
\affil[$\dagger$]{These authors contributed equally to this work.}

 \dates{}

 \doi{\url{https://doi.org/10.1364/OL.417032}}

\begin{abstract}
An ultrafast fiber chirped-pulse amplification laser system based on coherent combination of 16 ytterbium-doped rod-type amplifiers is presented. It generates 10~mJ pulse energy at 1~kW average power and 120~fs pulse duration. A partially helium-protected, two-staged chirped-pulse amplification grating compressor is implemented to maintain the close to diffraction-limited beam quality by avoiding nonlinear absorption in air.
\end{abstract}

\setboolean{displaycopyright}{true}

\begin{document}

\maketitle

\noindent Ultrafast lasers are highly demanded tools, allowing generation of high-harmonics \cite{Hadrich2016} or terahertz radiation \cite{Buldt2020} and opening up new and most challenging fields such as laser particle acceleration \cite{Leemans2011}. Besides necessarily requiring ultrashort pulse durations and high pulse energies at a diffraction limited beam quality, such applications typically strongly benefit from high average powers, improving processing or measurement speeds and signal-to-noise ratios. Altogether, reaching for the desired parameters demands the utmost of cutting-edge laser systems and pushes additional developments even further. 

For the generation of those parameters, especially three laser architectures based on ytterbium-doped active media come into question. While thin-disk and slab lasers especially shine in terms of high pulse energies \cite{Nubbemeyer2017, Kramer2020}, ultrafast fiber lasers prevail in terms of average power \cite{Mueller2020} and pulse duration, owing to their broad gain bandwidth. However, largely exploiting it requires well thought out spectral shaping to counteract gain narrowing. Alternatively or to reduce the pulse duration even further, different nonlinear pulse compression schemes have been developed. Noble gas-filled multi pass cells and capillaries, for instance, are widely used to reach the sub-$\SI{100}{fs}$ \cite{Kaumanns2018, Kramer2020, Grebing2020} or even few-cycle regime \cite{Nagy2019}. But these approaches have their limitations. For example, optically induced damage of the components and ionization of the gas so far allow only a part of the available performance to be compressed, while a finite compression factor limits the achievable output pulse duration and quality. Furthermore, such external nonlinear post-compression schemes pose a significant additional effort and typically require several meter long optical arrangements, becoming even larger with higher input pulse energies. Thus, a laser system approaching or even undercutting the $\SI{100}{fs}$ mark without an additional post-compression setup in a power and energy scalable way would be of great interest, tempting with lower overall complexity and simplified operability. Eventually, such a laser system could still be operated in combination with an additional nonlinear compression stage but, due to the significantly lower required compression factor, with potentially improved output pulse quality and way easier access to the few-cycle regime.

In this work, we present a high-power, high-energy fiber chirped-pulse amplification (CPA) laser system based on coherent beam combination \cite{Fan2005, Mueller2016} of 16 ytterbium-doped rod-type amplifiers. By applying aggressive but well-considered spectral shaping and exploiting the broad gain bandwidth of Yb-fiber amplifiers, the system generates by far the lowest CPA pulse duration at $\SI{1}{kW}$, multi-mJ performance to date. 

\begin{figure}[tb]
	\begin{center}
		\includegraphics[width=\linewidth*45/45]{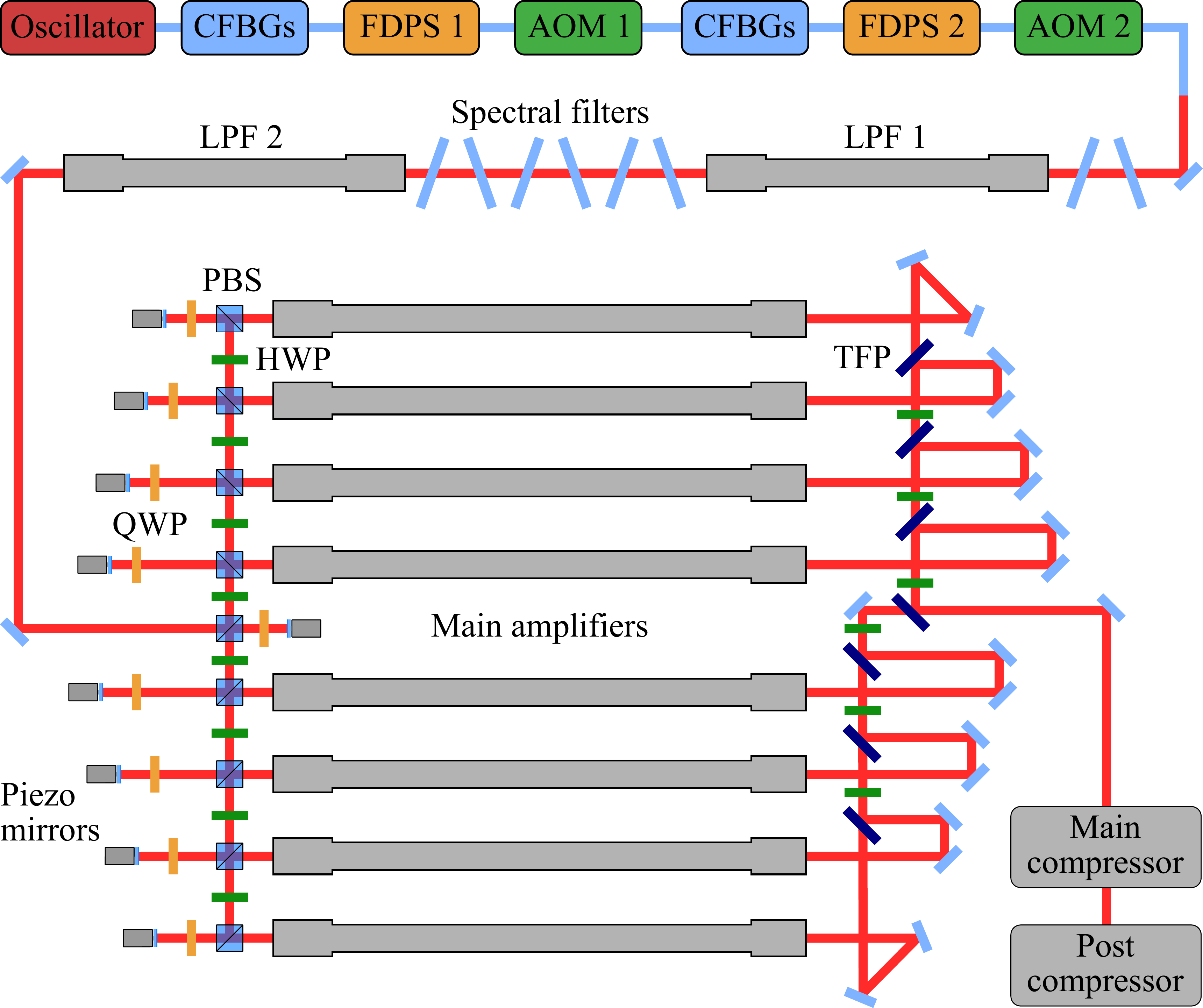}
		\caption{Schematic of the setup. For the sake of clarity, only the lower level of the main amplifier stage is depicted. However, the upper level looks similar. CFBG, chirped fiber Bragg gratings; FDPS, Fourier-domain pulse shaper; AOM, acousto-optic modulator; LPF, large-pitch fiber; QWP, quarter-wave plate; HWP, half-wave plate; PBS, polarizing beam splitter; TFP, thin-film polarizer.}
		\label{fig:setup}
	\end{center}
\end{figure}

A schematic of the system is depicted in Fig.~\ref{fig:setup}. The seed source is a broadband, ytterbium-based oscillator, emitting femtosecond pulses at a repetition rate of $\SI{64}{MHz}$. A completely fiber-integrated front end based on polarization maintaining $\SI{6}{\micro m}$-core fibers and components allows to condition the signal in terms of spectral amplitude, spectral phase and pulse repetition rate for the subsequent amplification to the high power regime. First, the pulses are pre-stretched to the nanosecond regime by a set of chirped fiber Bragg gratings (CFBG) with a hardcut of $\SI{40}{nm}$ centered at $\SI{1035}{nm}$, incorporating about one fourth of the total stretcher dispersion. This minimizes the accumulated nonlinear phase while still allowing to cleanly pick the pulse repetition rate without truncating the picked pulses, since the pre-stretched pulses of the oscillator pulse train are still separated sufficiently. The signal passes a Fourier-domain pulse shaper (FDPS~1), which gives full access to the spectral phase and amplitude of the transmitted signal and is later used to counteract gain narrowing and accumulated nonlinear phase. Next, the pulse repetition rate is reduced by a first acousto-optic modulator (AOM 1) and a second stage of CFBGs adds the remaining dispersion to the pulses. Another FDPS (FDPS~2) is passed in a double-pass configuration, allowing to apply even more spectral amplitude and phase shaping to the seed signal, if required. Finally, a second AOM (AOM 2) further reduces the pulse repetition rate, before the signal is coupled out of the fiber-integrated front end. In between, the signal is repeatedly re-amplified by core-pumped ytterbium-doped fiber amplifiers pumped at $\SI{976}{nm}$ wavelength.

The signal is free-space coupled into an ytterbium-doped large-pitch fiber (LPF) \cite{Limpert2012} amplifier which is, just like all following high-power amplifiers, mounted in a water-cooled aluminum module. To further compensate gain narrowing, a set of eight transmissive, free-space spectral notch filters is added to the system, two in front of the first LPF and six behind it. Each of the filters has a Gaussian-shaped spectral suppression with a full width at half maximum (FWHM) of about $\SI{6}{nm}$ and a peak attenuation of $\SI{1.5}{dB}$ at an angle-of-incidence-dependent wavelength around $\SI{1030}{nm}$. Finally, a second LPF pre-amplifier increases the average power to $\SI{25}{W}$. Using polarizing beam splitters (PBS) and half-wave plates (HWP), the beam is spatially split to seed the following main amplifier stage. It consists of 16 parallel ytterbium-doped distributed mode filtering fiber amplifiers with a mode-field diameter of about $\SI{62}{\micro m}$ and a length of $\SI{105}{cm}$, arranged in two layers of eight amplifiers each. 16 fiber-coupled pump diodes, each specified to deliver up to $\SI{250}{W}$ average power at a wavelength of $\SI{976}{nm}$, are used to counter-pump the main amplifiers individually. Thereafter, the single output beams are polarization beam combined using thin-film-polarizers (TFP). Anti-reflective coated laser windows after each combination step generate weak beam samples which are used for Hänsch-Couillaud polarization detection \cite{Hansch1980}. Together with corresponding piezo-mounted mirrors on the seed side of the main amplifiers, this enables interferometric phase stabilization and, thus, stable constructive interference.

The combined beam is recollimated to a diameter of $\SI{8}{mm}$ and sent through two subsequent free-space Treacy-type grating compressors with spectral hardcuts matched to the CFBGs. All gratings are highly efficient dielectric diffraction gratings with a line density of $\SI{1740}{l/mm}$. The first compressor consists of two gratings, a small one with $\SI{11}{cm}$ width and a $\SI{90}{cm}$ broad one, operated at an incidence angle of $\SI{60.5}{\degree}$. The distance between them, measured from center to center of the diffractive surfaces, is $\SI{155}{cm}$. The gratings are passed twice by the signal by means of a retro-reflecting roof mirror. Since this compressor removes most of the pulse's chirp but, however, not all of it, it is referred to as main compressor. The second compressor, named post compressor, consists of four dielectric gratings and is significantly smaller, but, in contrast to the main compressor, it is placed in an aluminum chamber filled with helium at a pressure of $\SI{1}{bar}$. The inert gas atmosphere is required since in air the pursued high peak and average power would lead to detrimental nonlinear absorption and thermal load \cite{Ali1983, Kartashov2006}, eventually spoiling the compressed beam beyond repair. To create a sufficiently pure and effective protective atmosphere, the chamber is evacuated and re-filled with helium to ambient pressure multiple times. Thus, the chamber needs to be vacuum compatible even if it is finally filled with helium at environmental pressure. Since the main compressor would require a vacuum chamber with dimensions of about $\SI{2.5}{m}$ by $\SI{1}{m}$ in a self-supporting structure, this would pose a challenging additional expense, potentially nevertheless leading to disastrous misalignment of the compressor during the pressure changes. By splitting the pulse compression into two stages, the requirement of a protective atmosphere could be limited to the second compression stage only, reducing the required dimensions of the helium chamber to practicable values of about $\SI{20}{cm}$ by $\SI{70}{cm}$. Finally, a sample of the fully compressed signal is taken for monitoring purposes using an output coupling mirror with 99\% reflectivity. Despite a total of eight passes over dielectric gratings, this compressor scheme has a high overall transmission of $86.5\%$, owing to the high diffraction efficiency of the gratings used. 

\begin{figure}[tb]
	\begin{center}
		\includegraphics[width=\linewidth*40/45]{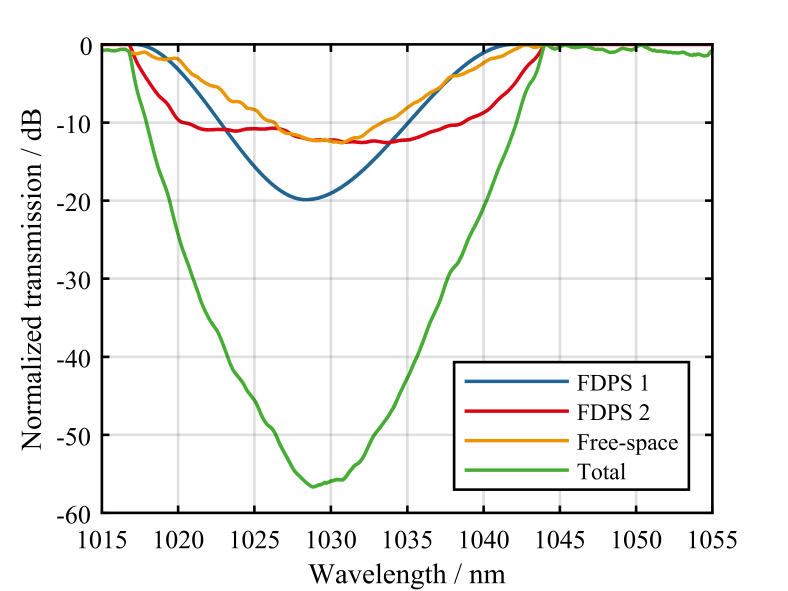}
		\caption{Single-pass amplitude mask settings of both FPDS in blue and red. Measured spectral transmission of all free-space notch filters in yellow. Total applied spectral shaping of the fiber front end (with FPDS 2 double-passed) and all free-space filters in green.}
		\label{fig:amplitude_shaping}
	\end{center}
\end{figure}

The aim of this experiment is to achieve a high average power of $\SI{1}{kW}$ and to maximize the pulse peak power at the output of the CPA system. Thus, a rather low pulse repetition rate of $\SI{100}{kHz}$ is chosen, where the fiber amplifiers work efficiently and still safely from any potential optically induced damage. The AOMs in the fiber front end are set accordingly. Furthermore, a broad spectrum is desired to minimize the output pulse duration. In addition, filling out the spectral hardcut as far as possible also maximizes the CPA-stretched pulse duration, reducing the accumulated nonlinear phase in the main amplifiers. To distribute the nonlinear phase homogeneously over the whole spectrum, facilitating the CPA-compression and the B-integral matching between the individual amplifiers, the shape of the amplified stretched pulses has to be close to rectangular. Knowing the theoretical dispersion values of the stretcher and compressor, the spectral shape required to generate such a rectangular stretched pulse is calculated. The output spectrum of a single channel is measured by an optical spectrum analyzer and, using the two FDPSs in the fiber frontend, adapted to fit the calculated spectrum as follows. FDPS~1 applies a spectral transmission mask with a Gaussian-like dip close to $\SI{1030}{nm}$ and a center attenuation of $\SI{20}{dB}$, see blue graph in Fig.~\ref{fig:amplitude_shaping}. This signal suppression around the gain maximum of the front end is mainly intended to disburden FDPS~2, since the attenuation in a single step has to be limited to avoid disruptive effects such as amplified spontanteous emission (ASE) in subsequent amplifiers. This approach is further enhanced by the eight free-space notch filters, whose total spectral transmission is depicted in yellow. Overall, this enables FPDS 2 to apply a broader but, with a maximum of $\SI{12.5}{dB}$ attenuation, rather shallow spectral suppression mask, since the gain peak has already mostly been taken care of. An automated computer algorithm then adapts the transmission mask of FDPS~2 such that the measured output spectrum is reshaped to the calculated form, eventually resulting in the desired rectangular stretched pulse profile in time-domain. The resulting transmission mask is colored red in Fig.~\ref{fig:amplitude_shaping}, the overall spectral shaping applied by both FDPSs (with the second one being double-passed) and all free-space notch filters is depicted in green and has a maximum attenuation of $\SI{57}{dB}$.

\begin{figure}[htb!]
	\begin{center}
		\includegraphics[width=\linewidth*40/45]{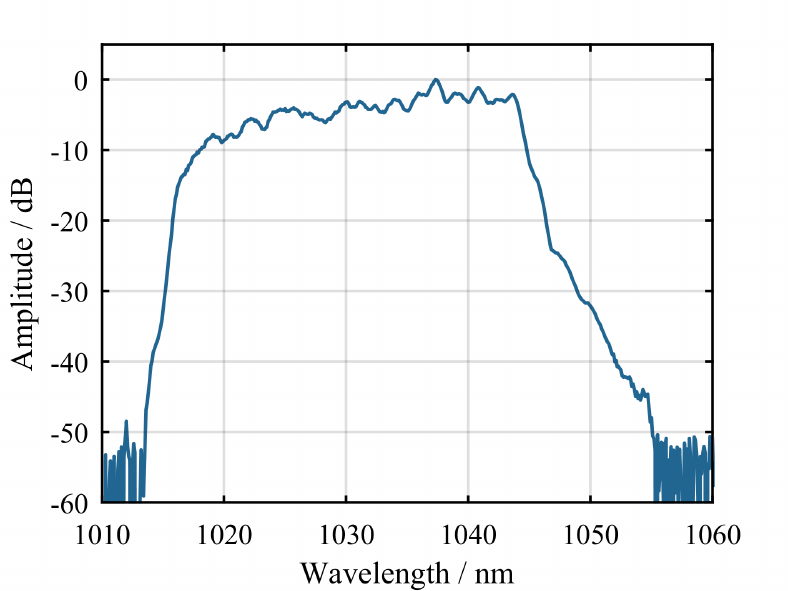}
		\caption{Spectrum of the combined and compressed $\SI{10}{mJ}$, $\SI{1}{kW}$ signal.}
		\label{fig:combined_spectrum}
	\end{center}
\end{figure}

The system is intended to allow reaching and investigating the fundamental limitations of currently available high-power, high-energy Yb-fiber amplifier technology, especially in terms of spectral restrictions posed by the gain bandwidth of the active medium. Thus, to eliminate potential technical limitations, the supported bandwidth of the front end and the CPA compressor was chosen to be generously broad. Although the system's spectral hardcut of $\SI{1015}{nm}$ to $\SI{1055}{nm}$ is still mostly filled out by the main amplifier output spectrum, as visible in Fig.~\ref{fig:combined_spectrum}, it is not notably limiting or truncating the optical bandwidth. However, the applied gain shaping is already very extensive. Forcing an even more excessive spectral shaping would not be beneficial, since the gain bandwidth of the main amplifiers is mostly exploited. Consequently, this would mainly reduce the overall amplifier efficiency and increase the ASE content in the output signal.

Still using a single amplifier channel, a part of the fully compressed beam sample taken at the end of the post compressor is used for the multi-photon intrapulse interference phase scan technique \cite{Lozovoy2004}. This allows to remove most of the remaining spectral phase on the compressed pulse, e.g. due to accumulated nonlinear phase or uncompensated higher order dispersion, by applying an appropriate inverse phase mask via FDPS~1 in the front end. 

The B-integrals of all main amplifiers have to be matched to achieve efficient coherent combination and to maintain the close to transform-limited compressed pulse. For this, the pump current of each amplifier is set individually to a value, where the auto-correlation (AC) traces and durations coincide, indicating similar nonlinear phases. Furthermore, the following combination optimization steps have to be done once. First, the optical path lengths of the individual main amplifier channels have to be matched using micrometer translation stages on the seed side of the amplifiers, placed underneath the piezo-driven mirrors. Next, the output signals of all channels have to be spatially superimposed. Using two cameras, one in the near field before the main compressor and one in the far field in the beam sample after the post compressor, allows simple and straight forward alignment of all 16 main amplifier channels. Finally, the rotation angles of the half-wave plates between the TFPs on the combining side are adjusted. 

\begin{figure}[htb!]
	\begin{center}
		\includegraphics[width=\linewidth*40/45]{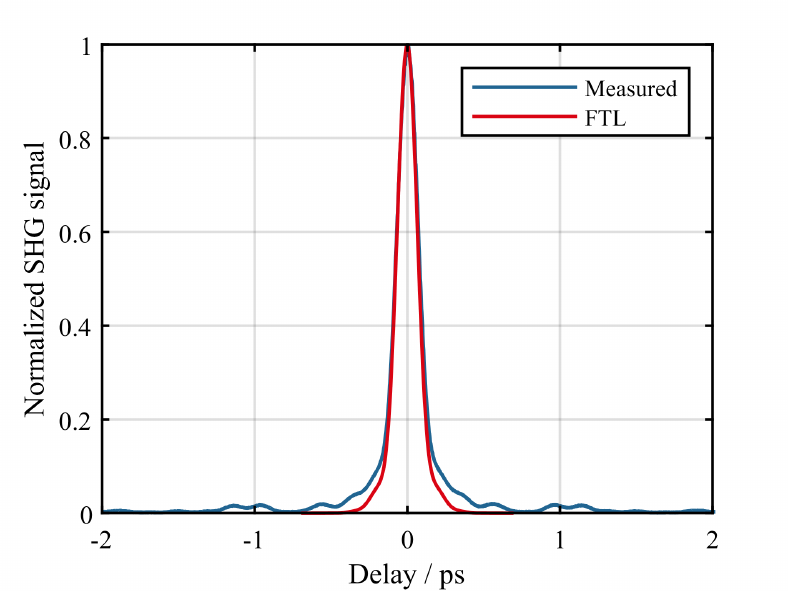}
		\caption{Depicted are the normalized AC signal of the combined, compressed $\SI{10}{mJ}$ pulses at $\SI{1}{kW}$ average power in blue and the Fourier transform-limited (FTL) pulse calculated from the corresponding spectrum (Fig.~\ref{fig:combined_spectrum}) in red. The corresponding pulse durations are $\SI{120}{fs}$ for the measured pulse and $\SI{115}{fs}$ for the FTL pulse.}
		\label{fig:combined_ac}
	\end{center}
\end{figure}

\begin{figure}[htb!]
	\begin{center}
		\includegraphics[width=\linewidth*40/45]{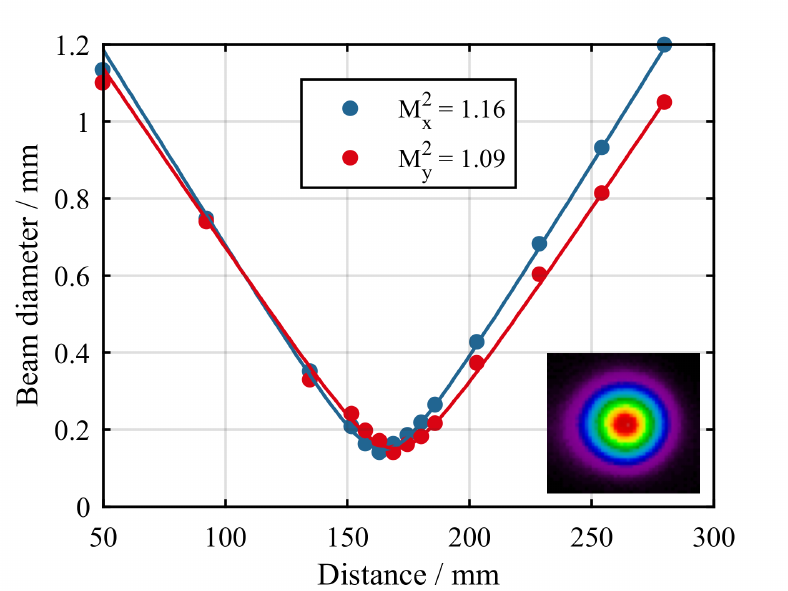}
		\caption{$\mathrm{M}^2$ measurement of the combined and compressed $\SI{10}{mJ}$, $\SI{1}{kW}$ beam. The intensity profile of the focus is shown in the inlay.}
		\label{fig:m_squared}
	\end{center}
\end{figure}

By optimizing these degrees of freedom once, a coherently combined and fully compressed average power of $\SI{1006}{W}$ is achieved, corresponding to $\SI{10}{mJ}$ pulse energy. The combining efficiency, defined as the ratio of the combined average power to the sum of all output powers from the individual main amplifiers, is 94\% at this operation point, indicating a good match of all previously mentioned parameters. The compressed pulse quality remains mostly unchanged by the combining, showing a similar, nearly transform-limited AC trace as depicted in Fig.~\ref{fig:combined_ac} with an FWHM AC duration of $\SI{164}{fs}$. The spectrum of the combined signal, shown in Fig.~\ref{fig:combined_spectrum} and having a $\SI{10}{dB}$-bandwidth of $\SI{27}{nm}$, is used to calculate the corresponding deconvolution factor, uncovering a compressed pulse duration of $\SI{120}{fs}$. Using the measured pulse duration and assuming an energy content of 90\% in the main feature, the pulse peak power is estimated to be $\SI{68}{GW}$. Additionally, the $M^2$ is measured using the $4\sigma$-method. It reveals a nearly diffraction limited beam quality with an $M^2 < 1.2$ in both axes as depicted in Fig.~\ref{fig:m_squared}. Even at these peak and average power figures, the close-to-ideal beam- and pulse quality are maintained during the propagation to the experimental site, confirming the effectiveness of the two-stage compressor approach.

In conclusion, a broadband high-power, high-energy ytterbium-doped fiber CPA laser system is presented, making extensive use of gain shaping in a wide-bandwidth-supporting, fiber-integrated front end. With the amplifier gain bandwidth being exploited as far as technically possible, a nearly transform-limited output pulse duration of $\SI{120}{fs}$ is achieved at $\SI{1}{kW}$ average power and $\SI{10}{mJ}$ pulse energy. Owing to a new two-staged grating compressor design involving a local protective helium atmosphere, the full performance can be delivered to experiments at close to diffraction-limited beam quality. While the achieved results already pose a record performance of fiber CPA systems, future experiments involving electro-optically controlled divided-pulse amplification \cite{Stark2019} and nonlinear post compression to the few cycle regime will boost peak power and pulse energy even further.

\section*{Acknowledgements} 
We thank Laseroptik GmbH for manufacturing the free-space spectral filters.

\section*{Funding}
European Research Council (835306); Fraunhofer-Gesellschaft (Cluster of Excellence "Advanced Photon Sources"); German Federal Ministry of Education and Research (funding program Photonics Research Germany, 13N15244).

\section*{Disclosures}
The authors declare no conflicts of interest.

\bibliography{references_10mJ.bib}

\bibliographyfullrefs{references_10mJ.bib}


\end{document}